# Observation of zero resistance above 100 K in $Pb_{10-x}Cu_x(PO_4)_6O$

Qiang Hou[1,2], Wei Wei[1,2], Xin Zhou[1,2], Yue Sun[1✉], Zhixiang Shi[1✉]

[1] Department of Physics, Southeast University, Nanjing 211189, China

**Abstract**：Room-temperature superconductivity has always been regarded as the ultimate goal in the fields of solid-state physics and materials science, with its realization holding revolutionary significance, capable of triggering significant changes in energy transmission and storage. However, achieving it poses various challenges. Recent research revealed that material $Pb_{10-x}Cu_x(PO_4)_6O$ displays room-temperature superconductivity under atmospheric pressure, sparking global interest in further exploration. Here, we utilized solid-phase synthesis to obtain a polycrystalline sample of $Pb_{10-x}Cu_x(PO_4)_6O$. X-ray diffraction confirmed its structural consistency with referenced literature. Zero resistance, which is an important evidence for superconductivity, was observed above 100 K under ambient pressure in our experiment. Our finding indicates that $Pb_{10-x}Cu_x(PO_4)_6O$ is a possible candidate for searching high-temperature superconductor.

**Introduction**

Superconductivity, first observed in 1911, has fascinated scientists due to its unique properties [1]. It has significantly impacted various fields, from fundamental physics research to practical applications like magnetic resonance imaging (MRI) machines and particle accelerators [2,3,4,5]. However, one defining limitation stood firmly in the way of its widespread implementation - the requirement of ultra-low temperatures to achieve the superconducting state. Therefore, over a century, lots of scientists have mad great efforts to search superconductors with higher transition temperature, $T_c$. The cuprate superconductors with $T_c$ over 150 K and iron-based superconductors with $T_c$ over 50 K have been discovered subsequently [6,7,8,9,10]. In 2015, $SH_3$ was reported to show $T_c$ ~203 K at a pressure of 155 GPa [11,12,13], then the superconductivity has been discovered in $YH_9$ and $LaH_{10}$ at 243 K and 260 K under ~ 200 GPa [14,15,16,17]. In March 2023, Gammon et al reported evidence of near ambient (1 GPa) superconductivity in lutetium-hydrogen-nitrogen, which attracts worldwide attention [18]. However, its reliability is still under debate [19]. Until now, the room temperature superconductor at ambient pressure has not been confirmed yet.

---

[1] Department of Physics, Southeast University, Nanjing 211189, China.

[2] These authors contributed equally: Qiang Hou, Wei Wei, Xin Zhou. ✉e-mail: sunyue@seu.edu.cn; zxshi@seu.edu.cn

Very recently, room-temperature superconductivity with $T_c \sim 400$ K at ambient pressure has been reported in Cu-doped lead-apatite $Pb_{10-x}Cu_x(PO_4)_6O$ ($0.9 < x < 1.1$) [20,21]. It is a very exited news for both the research community and the industry. However, more evidence for the superconductivity need to be confirmed. In this report, we successfully synthesized the $Pb_{10-x}Cu_x(PO_4)_6O$ compound, and observed the zero resistance above 100 K, which is an important evidence for superconductivity.

**Experiment**

In this manuscript, we synthesized polycrystalline sample of $Pb_{10-x}Cu_x(PO_4)_6O$ using solid-state sintering. The first step in this experiment involves synthesizing the precursor Lanarkite and $Cu_3P$. First of all, to obtain Lanarkite, a 1:1 stoichiometric mixture of high-purity $Pb(SO)_4$ powder (Aladdin 99.99%) and PbO powder (Aladdin 99.999%) was thoroughly mixed by grinding for one hour. Subsequently, the mixture was placed in two separate containers: a quartz tube under high vacuum and an alumina crucible exposed to air. The mixture was placed in a muffle furnace and heated to 725°C at a rate of 3 °C/min for 24 hours. We discovered that the mixture sintered in air ultimately transforms into phase $Pb_3O_2SO_4$, whereas the mixture sealed in a high-vacuum quartz tube transforms into phase Lanarkite $Pb_2(SO_4)O$. Specific analysis can be found in the next section. To synthesize $Cu_3P$, the following chemical equation is used: $3Cu + P \rightarrow Cu_3P$. Cu powder (Aladdin 99.9%) was thoroughly mixed with P (Aladdin 99.999%) in a molar ratio of 3:1 inside a Glovebox filled with argon. The mixture was sealed into high vacuum quartz tubes, slowly heated to 550 °C at a rate of 3 °C/min, and then kept in a furnace for 48 hours to obtain the precursor $Cu_3P$.

In order to synthesize $Pb_{10-x}Cu_x(PO_4)_6O$, we designed four different ratios for our experiment:

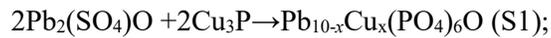
$2Pb_2(SO_4)O + 2Cu_3P \rightarrow Pb_{10-x}Cu_x(PO_4)_6O$ (S1);
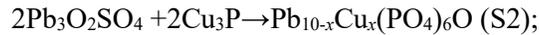
$2Pb_3O_2SO_4 + 2Cu_3P \rightarrow Pb_{10-x}Cu_x(PO_4)_6O$ (S2);
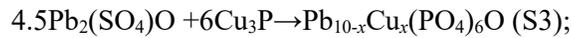
$4.5Pb_2(SO_4)O + 6Cu_3P \rightarrow Pb_{10-x}Cu_x(PO_4)_6O$ (S3);
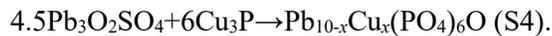
$4.5Pb_3O_2SO_4 + 6Cu_3P \rightarrow Pb_{10-x}Cu_x(PO_4)_6O$ (S4).

The batching and grinding processes are conducted inside a Glovebox filled with argon. Subsequently, the mixture was compressed into 8 mm cylindrical blocks at 6GPa and sealed in a high vacuum quartz tube. Finally, four evenly mixed raw materials were obtained. These uniformly mixed pellets were raised to a temperature of 925 °C at a rate of 1.5 °C/minute, held for 10 or 20 hours, and then cooled in a furnace to obtain our target product $Pb_{10-x}Cu_x(PO_4)_6O$.

The crystal structure of produced samples was examined using an X-ray diffractometer (XRD) with Cu-Kα radiation on a Bruker D8 Discover diffractometer. The in-plane electrical resistivity was carried out by the standard four-probe method on a physical property measurement system

(PPMS, Quantum Design). Magnetization measurement was preformed by the VSM (vibrating sample magnetometer) option of PPMS.

**Results and Discussion**

Figure 1 shows the XRD patterns of the precursors $Cu_3P$ and $Pb_2(SO_4)O$, along with all the synthesized products. The data for $Cu_3P$ in Fig. 1(a) is in complete agreement with the standard PDF card from software jade 6.5, indicating the pure phase without impurities. For the precursor $Pb_2(SO_4)O$, the formation of $Pb_2(SO_4)O$ and $Pb_3O_2SO_4$ were observed under sintering mixture of $Pb(SO)_4$ and PbO in air and vacuum, respectively. To examine the impact of precursor ratios on the final product, we prepared four samples, namely S1, S2, S3, and S4, each with different precursors and ratios, as detailed in Table 1. In Fig. 1(c), the powder XRD patterns of S1 and S2 exhibit identical peak positions, as do those of S3 and S4, suggesting that the choice of precursor, $Pb_2(SO_4)O$ or $Pb_3O_2SO_4$, has limited effect on the target product. Instead, the peaks of S1 and S3 (or S2 and S4) are almost completely different, indicating that the ratio of precursors $Pb_2(SO_4)O$: $Cu_3P$ (or $Pb_3O_2SO_4 : Cu_3P$) is the key to the synthesis of samples.

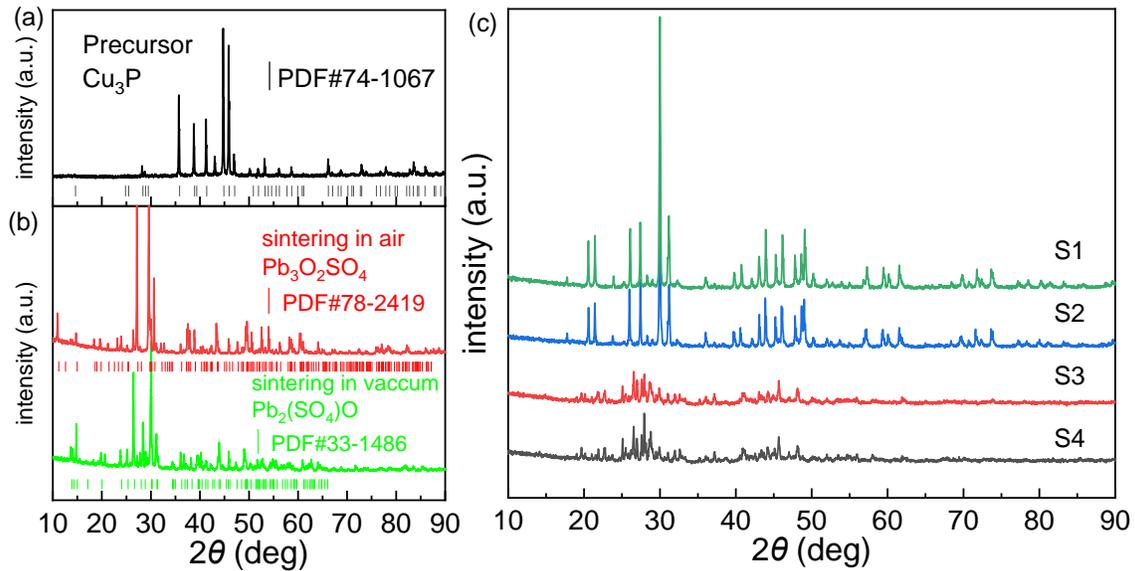

Fig. 1 X-ray diffraction patterns of (a) precursor $Cu_3P$, (b) $Pb_2(SO_4)O$ (sintering in vacuum), $Pb_3O_2SO_4$ (sintering in air), and (c) target product of $Pb_{10-x}Cu_x(PO_4)_6O$.

**TABLE 1. Precursors and ratio of four samples.**

|  | 1: 1 | 4.5: 6 |
|---|---|---|
| $Pb_2(SO_4)O: Cu_3P$ | S1 | S3 |
| $Pb_3O_2SO_4 : Cu_3P$ | S2 | S4 |

To validate the successful synthesis of $Pb_{10-x}Cu_x(PO_4)_6O$, Figure 2 displays the XRD patterns of the sample in ref. [21] and S1. Remarkably, the XRD diffraction peaks of sample S1 are consistent with the ref. [21], and on this basis, the phase formation has been improved, with no distinct $CuS_2$ peak observed. It provides a strong evidence for the successful synthesis of $Pb_{10-x}Cu_x(PO_4)_6O$. Moreover, high-quality samples guarantee the accuracy of the results we obtained.

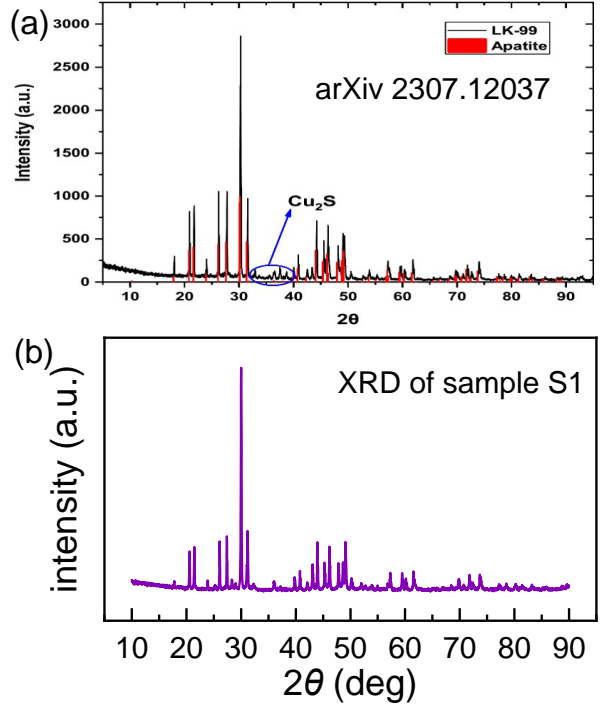

Fig. 2 X-ray diffraction patterns of (a) ref. [21] and (b) sample S1.

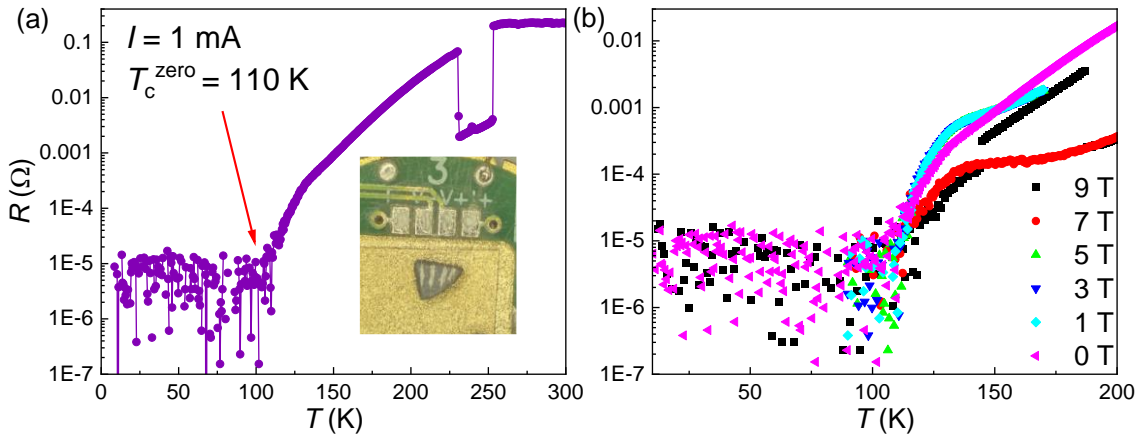

Fig 3. (a) Temperature dependence of resistance for sample S1 pallet; (b) Temperature dependence of resistance for sample S1 under magnetic fields from 0 to 9 T.

In order to further investigate the physical properties of $Pb_{10-x}Cu_x(PO_4)_6O$, we conducted electrical measurements on sample S1, which was divided into several small pieces. Due to the fragility of the samples, it is difficult to cut them into regular shape. Therefore, we selected some small pieces for measurement, as shown in the insert of Figure 3(a). Fig. 3(a) shows the temperature dependence of resistance sample S1, with a clear zero resistance characteristic observed at 110 K, indicating the possibility of superconductivity. It is worth noting that some other pieces exhibit semiconductor behavior, consistent with the literature [22]. Above 250K, two significant resistance jumps are clearly observed, and the temperature-rise and cooling measurement curves perfectly coincide. It is not clear what is the cause of these large jumps, which may be due to the influence of the electrode contact. We also measured the temperature dependence of resistance under fields ranging from 0 to 9 T, as shown in Fig. 3 (b). Inset shows the image of the measured sample S1 with electrodes attached. As the applied magnetic field increases to 5 T, $T_c^{onset}$ is gradually suppressed, but $T_c^{zero}$ still reaches more than 110 K. When the magnetic field is increased to more than 7 T, the resistance trend is abnormal, which is contrary to normal superconductors that the external magnetic field will simply reduce the critical temperature of superconductivity. The specific reason needs our further study. More importantly, the zero resistance is robust under magnetic field, which indicates possible large upper critical field in this material. To further verify the superconducting properties, we conducted magnetic measurements on the sample, but unfortunately, no obvious Meissner signal was observed, indicating that the superconducting volume fraction of the sample may be very small. The preparation of high-purity samples are still a challenging task.

**Conclusion**

In conclusion, we successfully synthesized the compound $Pb_{10-x}Cu_x(PO_4)_6O$, and observed the zero resistance above 100 K. However, the Meissner effect has not been observed yet in our samples, which suggests that the superconducting volume is relatively low. We still need more evidences to confirm the superconductivity, and to identify which component is in charge of the zero resistance (superconductivity). Besides, whether the $T_c$ could be enhanced up to room temperature is still an open question.

**Acknowledge**

This work was partly supported by the National Key R&D Program of China (Grant No. 2018YFA0704300), the Strategic Priority Research Program (B) of the Chinese Academy of Sciences (Grant No. XDB25000000), and the National Natural Science Foundation of China (Grant No. U1932217).